\documentstyle[amssymb,aps,12pt]{revtex}

\begin{document}
\draft
\author{O. B. Zaslavskii}
\address{Department of Mechanics and Mathematics, Kharkov V.N. Karazin's National\\
University, Svoboda\\
Sq.4, Kharkov 61077, Ukraine\\
E-mail: ozaslav@kharkov.ua}
\title{Two-dimensional quantum-corrected black hole in a finite size cavity}
\maketitle

\begin{abstract}
We consider the gravitation-dilaton theory (not necessarily exactly
solvable), whose potentials represent a generic linear combination of an
exponential and linear functions of the dilaton. A black hole, arising in
such theories, is supposed to be enclosed in a cavity, where it attains
thermal equilibrium, whereas outside the cavity the field is in the Boulware
state. We calculate quantum corrections to the Hawking temperature $T_{H}$,
with the contribution from the boundary taken into account. Vacuum
polarization outside the shell tend to cool the system. We find that, for
the shell to be in the thermal equilibrium, it cannot be placed too close to
the horizon. The quantum corrections to the mass due to vacuum polarization
vanish in spite of non-zero quantum stresses. We discuss also the canonical
boundary conditions and show that accounting for the finiteness of the
system plays a crucial role in some theories (e.g., CGHS), where it enables
to define the stable canonical ensemble, whereas consideration in an
infinite space would predict instability.
\end{abstract}



\section{Introduction}

Two-dimensional dilaton (2D) gravity serves as an excellent tool for
studying (at least, on the semiclassical level) quantum effects in
gravitation and constructing the prototype of (yet unfinished) 4D quantum
gravity. In the first place, it concerns black hole physics, where
string-inspired models \cite{callan} became very popular during last decade
and enabled to trace in the simplified context many effects, typical of 4D
black hole physics, such as black hole evaporation, thermodynamic features,
etc. (for reviews see, e.g., recent papers \cite{dv}, \cite{od}). On the
other hand, studies in 2D black hole physics revealed the fact that, in some
aspects, such theories look rather unusual and open new interesting
possibilities, absent in general relativity and deserving treatment on their
own. For example, the Hawking temperature $T_{H}$ in the classical CGHS
model and some its semiclassical generalizations is a constant, not
depending on the horizon radius. This makes the question of black hole
thermodynamics, which is one of the most important black hole features,
quite non-trivial. First, as pure classical thermodynamics is poor for such
systems, quantum backreaction and corresponding quantum corrections to the
Hawking temperature become crucial for the calculation of the heat capacity.
Second, even with quantum backreaction taken into account, some models, that
include a wide family of exactly solvable ones, still exhibit no quantum
corrections to $T_{H}$. To obtain substantial thermodynamics, one should
taken into account that for self-gravitating systems a finite size can be
crucial in careful constructing the canonical ensemble \cite{york86}. We
will see below that the competition of these two factors - small quantum
corrections and large (but finite) spatial size may lead to well-defined
thermal properties even in situations, when the pure classical approach
gives no sensible answer.

The issue of the finiteness of the system has one more aspect. Consider a
black hole enclosed inside a reflecting shell (that, in 1+1 case, represents
a point) in thermal equilibrium with its Hawking radiation. If the shell is
perfect, all Hawking radiation is concentrated inside and no radiation comes
outside. However, the entire object ''black hole+radiation+shell'' curves
spacetime and serves as a source of gravitation field outside. In turn, this
leads to the appearance of quantum stresses even in an otherwise empty
space. In other words, the field state is supposed to be the Hartle-Hawking
inside the shell and the Boulware one outside. Usually these states are
opposed in the 4D world, where the first state is attributed to a black
hole, while the second one corresponds to a relativistic star. However, the
boundary effects may lead to their overlap and, thus, the Boulware state
becomes relevant for black hole thermodynamics, so this effect deserves
attention on its own.

The quantum corrections to $T_{H}$ were calculated in \cite{pl} for the
black hole in the Hartle-Hawking state for the particular case of the CGHS
model, but the contribution of vacuum polarization was neglected there (that
looks quite reasonable, if a boundary is situated sufficiently far from the
horizon). Recently, these corrections were considered in \cite{pos} (in the
quite different approach) for a slowly evaporating black hole in the Unruh
state. The results differ by the sign that seem to affect the sign of the
heat capacity in an infinite space. This prompts us to consider the issue of
stability carefully, with proper account for the finite size of the system.
We will see that the stable canonical ensemble can be defined even in the
cases when consideration in an infinite space would give the negative heat
capacity.

The paper is organized as follows. In Sec. II we list basic equations,
governing the gravitational-dilatonic system with minimal fields, and, by
considering quantum backreaction as perturbation, derive the quantum
corrections to the Hawking temperature in an infinite space. This
generalizes our previous result which was obtained for the particular case
of the CGSH model. In Sec. III we consider a black hole, enclosed inside a
perfect reflecting shell, outside of which the quantum field in the Boulwars
state, heated to some temperature. For exactly solvable models we find the
modified Hawking temperature exactly, for a generic model we find the main
quantum corrections. In Sec. IV we analyze the canonical ensemble and its
stability with account for both quantum backreaction and finiteness of the
system. We also discuss briefly the case of the microcanonical ensemble. In
Sec. V we summarize main results.

\section{Quantum corrections to Hawking temperature in an infinite space}

Let us consider the system governed by the action 
\begin{equation}
I=I_{gd}+I_{PL}\text{, }  \label{1}
\end{equation}
where gravitation-dilaton part 
\begin{equation}
I_{gd}=\frac{1}{2\pi }\int_{M}d^{2}x\sqrt{-g}[F(\phi )R+V(\phi )(\nabla \phi
)^{2}+U(\phi )]\text{,}  \label{2}
\end{equation}
$I_{PL\text{ }}$is the Polyakov-Liouville action incorporating effects of
Hawking radiation of minimal fields and its backreaction on spacetime for a
multiplet of N conformal scalar fields (we omit boundary terms in the
action). As is known, it can be written down in the form 
\begin{equation}
I_{PL}=-\frac{\kappa }{2\pi }\int_{M}d^{2}x\sqrt{-g}[\frac{(\nabla \psi )^{2}%
}{2}+\psi R]\text{,}  \label{3}
\end{equation}
$\kappa =\frac{
\rlap{\protect\rule[1.1ex]{.325em}{.1ex}}h%
N}{24}$. The function $\psi $ obeys the equation 
\begin{equation}
\square \psi =R\text{.}  \label{r}
\end{equation}
Varying the action with respect to the metric, we get 
\begin{equation}
\delta I=\frac{1}{4\pi }\int d^{2}x\sqrt{-g}G_{\mu \nu }\delta g^{\mu \nu
}=0.
\end{equation}

For static spacetimes (with which we are dealing with in this paper) in the
Schwarzschild gauge 
\begin{equation}
ds^{2}=-fdt^{2}+f^{-1}dx^{2}  \label{gauge}
\end{equation}
field equations take the following explicit form (see, for example, eqs.
(23) and (24) of \cite{ee}): 
\begin{equation}
G_{0}^{0}=2f\frac{\partial ^{2}\tilde{F}}{\partial x^{2}}+\frac{\partial f}{%
\partial x}\frac{\partial \tilde{F}}{\partial x}-U-\tilde{V}f\left( \frac{%
\partial \phi }{\partial x}\right) ^{2}=0\text{,}  \label{sc00}
\end{equation}

\begin{equation}
G_{1}^{1}=\frac{\partial f}{\partial x}\frac{\partial \tilde{F}}{\partial x}%
-U+\tilde{V}f\left( \frac{\partial \phi }{\partial x}\right) ^{2}=0\text{.}
\label{sc11}
\end{equation}

Here 
\begin{equation}
\tilde{F}=F-\kappa \psi \text{, }\tilde{V}=V-\frac{\kappa }{2}\left( \frac{%
d\psi }{d\phi }\right) ^{2}\text{.}
\end{equation}

In what follows we will use notations $U=4\lambda ^{2}u$ for the potential
and $z=\lambda x$ for a coordinate. It is also convenient to take the sum
and difference of Eqs. (\ref{sc00}) and (\ref{sc11}) that gives us 
\begin{equation}
\lbrack \frac{\partial ^{2}\tilde{F}}{\partial \phi ^{2}}-\tilde{V}]\left( 
\frac{\partial \phi }{\partial z}\right) ^{2}+\frac{\partial \tilde{F}}{%
\partial \phi }\frac{\partial ^{2}\phi }{\partial z^{2}}=0\text{,}
\label{vx}
\end{equation}
\begin{equation}
4u=(f\frac{\partial \tilde{F}}{\partial \phi }\frac{\partial \phi }{\partial
z})_{,\phi }\frac{\partial \phi }{\partial z}\text{.}  \label{uf}
\end{equation}

In the conformal frame 
\begin{equation}
ds^{2}=f(-dt^{2}+d\sigma ^{2})  \label{con}
\end{equation}
we have ($y\equiv \lambda \sigma $) 
\begin{equation}
\lbrack \frac{\partial ^{2}\tilde{F}}{\partial \phi ^{2}}-\tilde{V}-f^{-1}%
\frac{\partial f}{\partial \phi }]\left( \frac{\partial \phi }{\partial y}%
\right) ^{2}+\frac{\partial \tilde{F}}{\partial \phi }\frac{\partial
^{2}\phi }{\partial x^{2}}=0  \label{vy1}
\end{equation}

and 
\begin{equation}
4u=f^{-1}\frac{\partial ^{2}\tilde{F}}{\partial y^{2}}\text{.}  \label{u}
\end{equation}

In what follows we will dwell upon the string-inspired models of the form

\begin{equation}
F=\exp (-2\phi )+b\kappa \phi \text{, }V=4\exp (-2\phi )+c\kappa \text{, }%
u=\exp (-2\phi )\text{.}  \label{fv}
\end{equation}
Then the solutions of Eqs. (\ref{r}), (\ref{vx}) and (\ref{uf}) (or (\ref
{vy1}), (\ref{u})), regular on the horizon of a black hole, in the main
approximation with respect to $\kappa $ look like

\begin{equation}
\psi =-2\phi +O(\kappa )\text{,}
\end{equation}
\begin{equation}
z=-\phi +\frac{\kappa }{4}e^{2\phi }(1-\frac{c}{2})\text{,}  \label{z}
\end{equation}
\begin{equation}
f=1-a\exp (2\phi )+\kappa \exp (2\phi )\{\frac{q}{2}[1-a\exp (2\phi )]+(1-%
\frac{c}{2})\phi \}.  \label{fn}
\end{equation}
\begin{equation}
a=\exp (-2\phi _{+})+\kappa \phi _{+}(1-\frac{c}{2})\text{.}
\end{equation}
\begin{equation}
T_{H}=T_{0}[1+\frac{\kappa }{2}\exp (2\phi _{+})q]\text{, }T_{0}\equiv \frac{%
\lambda }{2\pi }\text{, }  \label{t}
\end{equation}

where $q\equiv b+\frac{c}{2}+1$. It is seen from (\ref{t}) that,
classically, the Hawking temperature is a constant and all dependence on the
horizon position arises only via quantum corrections. For the CGHS model $%
c=0=b$, $q=1$ and we return to the result, that can be obtained by the
limiting transition $\phi _{B}\rightarrow -\infty $ from eq. (18) of Ref. 
\cite{pl}. If 
\begin{equation}
b=2(d-1),c=2(1-2d)\text{,}  \label{es}
\end{equation}
that is equivalent to 
\begin{equation}
q=b+\frac{c}{2}+1=0\text{,}  \label{esq}
\end{equation}
the model becomes exactly solvable \cite{cruz} and reduces, in particular
cases, to the RST ($c=0$, $b=-1$) \cite{rst} or BPP ($c=2$, $b=-2$) \cite
{bose} ones. In this case quantum corrections to the Hawking temperature
vanish in accordance with observations, made in \cite{sol96} and \cite{exact}%
.

\section{Matching Hartle-Hawking and Boulware states and role of boundary}

In the 1+1 world, a ''shell'' represents a point. If it is present, field
equations modify to 
\begin{equation}
G_{\mu }^{\nu }=S_{\mu }^{\nu }\text{,}
\end{equation}
where $S_{\mu }^{\nu }$ is a dimensionless stress-energy of the shell,
containing only delta-like terms. We assume that quantities $y$, $f$, $\phi $%
, $\tilde{F}$ are continuous across the shell, while first derivatives $%
\frac{\partial \tilde{F}}{\partial y}$may experience jumps. Then it follows
from explicit expressions (\ref{sc00}), (\ref{sc11}) or (\ref{vx}), (\ref{uf}%
) that $G_{1}^{1}$ is bounded across the shell, so $S_{1}^{1}=0$, whereas $%
G_{0}^{0}$ may contain delta-like singularities: 
\begin{equation}
S_{0}^{0}=-m\delta ^{(1)}(y-y_{B})\text{, }m\equiv -2[\left( \frac{\partial 
\tilde{F}}{\partial y}\right) _{+}-\left( \frac{\partial \tilde{F}}{\partial
y}\right) _{-}]\text{,}  \label{mf}
\end{equation}
where the parameter $m$ can be regarded as the mass of the shell, $y_{B}$ is
its position, ''+''\ or ''-'' means ''$y_{B}+0$'' and ''$y_{B}-0$'',
respectively. The delta function $\delta ^{(1)}$ is normalized according to 
\begin{equation}
\int dy\sqrt{g}\delta ^{(1)}(y-y_{B})=1\text{,}
\end{equation}
where the index''B'' refers to the boundary.

\subsection{Exactly solvable case}

First, we consider the exactly solvable case, when the coefficients obey the
relationship (\ref{es}). As is shown in \cite{exact}, \cite{thr}, the metric
function, describing a black hole in an infinite space, is equal in this
case to

\begin{equation}
f=\exp (2\phi +2y).  \label{f}
\end{equation}
The coefficient in from of $f$ is chosen in (\ref{f}) in such a way, that $%
f\rightarrow 1$ at right infinity, where the spacetime is flat.

In so doing, 
\begin{equation}
\tilde{F}=\exp (2y)+\tilde{F}_{+}\text{,}  \label{fy}
\end{equation}
the index ''+'' refers to the horizon, which is situated at $y=-\infty $, so 
\begin{equation}
f=\exp (2\phi )(\tilde{F}-\tilde{F}_{+})\text{.}  \label{ff}
\end{equation}

Let now a 0-dimensional point-like perfect shell between the horizon and
right infinity be situated at some $y_{B}$. To the left to the shell, the
field is in the Hartle-Hawking state, while to the right to the shell it is
in the Boulware state. Consider the solution in both region separately and,
afterwards, sew them at $y=y_{B}$.

First, consider the region $y<y_{B}$. We can exploit the already obtained
solution (\ref{f}) but with the reservation that there is a freedom in the
choice of the conformal coordinate that preserves the conformal gauge (\ref
{con}). The coordinate $y$ can be rescaled as $y\rightarrow Ay$, where $A$
is a constant. Apart from this, there is also a freedom in translations $%
y\rightarrow y+const$. For the solution in an infinite space it was
inessential since, due to the condition $f(\infty )=1$, it had to be reduced
to (\ref{f}). However, now there is no right infinity in the left region and
such parameters should be kept arbitrary, their values will be fixed from
matching the solutions in two regions (see below). Therefore, now we should
write

\begin{equation}
\tilde{F}=\tilde{F}^{(0)}=\alpha \exp (2Ay)+\tilde{F}_{+}\text{,}
\end{equation}
\begin{equation}
f_{l}=\frac{\alpha A^{2}}{u}\exp (2Ay)=\frac{A^{2}}{u}(\tilde{F}-\tilde{F}%
_{+})\text{,}
\end{equation}
where $\alpha $ and $A$ are constants. The Hawking temperature for the
metric (\ref{con}) $T_{H}=\frac{1}{4\pi }\lim_{y\rightarrow -\infty }f^{-1}%
\frac{df}{dy}$, so 
\begin{equation}
T_{H}=\frac{\lambda }{2\pi }A\text{.}
\end{equation}
Consider now the region to the right to shell, $y>y_{B}$. Now we should take
into account that the function is determined from eq. (\ref{r}) up to the
solution of the homogeneous equation that is proportional to $y$. Again, we
may exploit the solution in an infinite spacetime, obtained in \cite{thr}:

\begin{equation}
\psi =\psi _{0}+\frac{\gamma }{\lambda }y\text{.}
\end{equation}
\begin{equation}
\tilde{F}=\tilde{F}^{(0)}-\kappa \frac{\gamma }{\lambda }y\text{,}
\end{equation}
where $\tilde{F}^{(0)}=F-\kappa \psi _{0}$, $\psi _{0}$ is bounded on the
horizon (for the exactly solvable under discussion $\psi _{0}=-2\phi $), 
\begin{equation}
\tilde{F}^{(0)}=\exp (2y)-By+E\text{,}  \label{fc}
\end{equation}
where $E$ is a constant, 
\begin{equation}
B=\kappa (1-\frac{T^{2}}{T_{0}^{2}})\text{.}
\end{equation}
\begin{equation}
f_{r}=\frac{e^{2y}}{u}\text{.}
\end{equation}
Here $T$ is the temperature of the thermal gas at the right infinity. As is
shown in \cite{thr}, the constant $\gamma $ is connected with $T$ according
to 
\begin{equation}
\gamma =2\lambda (\frac{T}{T_{0}}-1)\text{.}  \label{ga}
\end{equation}

The attempt of applying the above formulas to the region near the black hole
horizon ($y\rightarrow -\infty $) shows that the quantity $\psi $ diverges
there and so does the Polyakov-Liouville-stresses \cite{found}. However,
this problem does not arise now since the region, in which these formulas
are valid, is restricted by the condition $y>y_{B}$ and does not include the
horizon.

On the boundary $f_{l}(\phi _{B})=f_{r}(\phi _{B})$, whence

\begin{equation}
\alpha =A^{-2}\exp [2y_{B}(1-A)]\text{.}
\end{equation}
\begin{equation}
E=(\tilde{F}_{B}^{(0)}-\tilde{F}_{+})(1-A^{2})+By_{B}+\tilde{F}_{+}\text{.}
\end{equation}
Calculating the difference $[\left( \frac{\partial \tilde{F}}{\partial y}%
\right) _{+}-\left( \frac{\partial \tilde{F}}{\partial y}\right) _{-}]$ and
remembering eq. (\ref{mf}), we obtain 
\begin{equation}
2\exp (2y_{B})[1-\frac{1}{A}]-\tilde{B}+-\frac{m}{2}=0\text{,}
\end{equation}
\begin{equation}
\tilde{B}=B+\frac{\gamma }{\lambda }\kappa =-\frac{\kappa \gamma ^{2}}{%
4\lambda ^{2}}=-\kappa (1-\frac{T}{T_{0}})^{2}\text{.}
\end{equation}
Substituting $\exp (2y_{B})=A^{2}(\tilde{F}_{B}-\tilde{F}_{+})$, we obtain
the equation 
\begin{equation}
A^{2}-A-\frac{\tilde{B}-m/2}{2(\tilde{F}_{B}-\tilde{F}_{+})}=0\text{,}
\end{equation}
\begin{equation}
A=\frac{1}{2}\left( 1+\sqrt{1-\frac{2\left| \tilde{B}\right| +m}{(\tilde{F}%
_{B}-\tilde{F}_{+})}}\right) \text{, }\tilde{B}=-\left| \tilde{B}\right| 
\text{.}  \label{a}
\end{equation}

We choose the root of the quadratic equation for which $A=1$, when $\kappa
=0=m$.

It follows from (\ref{a}) that A$_{\min }=\frac{1}{2}$, when $m=\tilde{F}%
_{B}-\tilde{F}_{+}+2\tilde{B}$. If $m>0$, $\frac{1}{2}<A<1$. Quantum effects
for the temperature are compensated by the shell mass if $m=2\tilde{B}<0$.
In the limit $y_{B}\rightarrow \infty $, when $\tilde{F}_{B}\rightarrow
\infty $, quantum corrections tend to zero: $\frac{\Delta T_{H}}{T}\simeq -%
\frac{1}{2}\frac{\left| \tilde{B}\right| +m/2}{\tilde{F}_{B}-\tilde{F}_{+}}$%
. It is seen from Eq. (\ref{a}) that both the quantum effects and the shell
with a positive mass tend to cool a system.

From the expression (\ref{a}) it follows the restriction on the position of
the shell that cannot be placed too close to the horizon, if we want to
maintain thermal equilibrium inside the shell and the Boulware state, heated
to the temperature $T$, outside:

\begin{equation}
\tilde{F}_{B}-\tilde{F}_{+}-2\left| \tilde{B}\right| -m>0\text{.}
\label{res}
\end{equation}

As the Riemann curvature $R=-\frac{\lambda ^{2}}{f}\frac{d^{2}\ln f}{dy^{2}}$%
, it follows from (\ref{f}) and (\ref{mf}) that for the exactly solvable
models the delta-like part of the curvature is equal to $R_{s}=\frac{%
m\lambda ^{2}}{\tilde{F}^{\prime }(\phi _{B})}\delta ^{(1)}(y-y_{B})$%
.Therefore, for a massless shell the geometry is smooth across the shell.

\subsection{Generic case. Perturbative approach}

Now let the system be of the type (\ref{fv}) with generic coefficients, not
necessarily obeying the condition of exact solvability (\ref{esq}). Then the
explicit formulas can be obtained perturbatively in $\kappa $, matching the
solutions to the left and to the right from the shell, following the same
line, as in the previous case. In so doing, we retain only terms of the zero
and first order in $\kappa $. Omitting details of calculations, which are
rather straightforward, we list only basic formulas. To the right from the
shell the relationship between the dilaton and spatial coordinate reads 
\begin{equation}
\frac{d\phi }{dz}=-1-\kappa \frac{\exp (2\phi )}{2}[1-\frac{c}{2}-\frac{D}{%
1-k\exp (2\phi )}]\text{,}  \label{dz}
\end{equation}
the metric function has the form

\begin{equation}
f=1-k\exp (2\phi )+\kappa \exp (2\phi )\{\frac{q}{2}[1-k\exp (2\phi )]+(1-%
\frac{c}{2}-D)\phi +\frac{D}{2}\ln (1-k\exp (2\phi )\}\text{,}  \label{ri}
\end{equation}

\begin{equation}
D=-\frac{\gamma }{\lambda }(1+\frac{\gamma }{4\lambda })=1-\frac{T^{2}}{%
T_{0}^{2}}\text{.}  \label{d}
\end{equation}
In the exactly solvable case $q=0$, 
\begin{equation}
f=1-k\exp (2\phi )+\kappa [(b+2-D)\phi \exp (2\phi )+\frac{D}{2}\exp (2\phi
)\ln (1-ke^{2\phi })]\text{,}
\end{equation}

that can be also obtained directly from (\ref{fc}). If $D=0$, eq (\ref{fn})
is reproduced.

To the left from the shell $f=A^{2}\tilde{f}$, where $\tilde{f}$ is given by
eq. (\ref{fn}).

Matching the solution in two region, we obtain from (\ref{mf})

\begin{equation}
A=\frac{1}{2}(1+\sqrt{1-\frac{m+2\kappa \left| \tilde{D}\right| }{Q})}\text{%
, }Q=[\exp (-2\phi _{B})-\exp (-2\phi _{+})]+\frac{\kappa }{2}(1-\frac{c}{2}%
)(\phi _{B}-\phi _{+})\text{, }\left| \tilde{D}\right| =(1-\frac{T}{T_{0}}%
)^{2}\text{.}
\end{equation}

For the massless shell, with the same accuracy (with terms $\kappa ^{2}$ and
higher discarded) 
\begin{equation}
A=1-\frac{\kappa \left| \tilde{D}\right| }{2Q}\text{,}
\end{equation}
\begin{equation}
T_{H}=T_{0}(1+\kappa \varepsilon )\text{, }\varepsilon =\frac{q}{2}\exp
(2\phi _{+})-\frac{\left| \tilde{D}\right| }{2Q}\text{.}  \label{td}
\end{equation}

For the CGSH model $b=0=c$, $q=1$, for $T=0\,$($\left| \tilde{D}\right| =1$)
in the limit $\left| \phi _{B}\right| \gg \left| \phi _{+}\right| $,
neglecting the term $\exp (-2\phi _{+})$ in the denominator, we obtain that $%
\varepsilon =\frac{q}{2}\exp (2\phi _{+})-\frac{1}{2}\exp (-2\phi _{B})$
that coincides with eq. (18) of \cite{pl}\footnote{%
It was stated in Ref. \cite{pl} that the shell should be inevitably massive
to maintain equilibrium, whereas in the present paper we mention a massless
shell ($m=0$), while comparing the results. There is no contradiction here
since these statements refer to different quantities. It follows from (\ref
{mf}) that the mass is linear functional of the quantity $\tilde{F}=F-\kappa
\psi $ and, correspondingly, can be spited in two parts - $m_{F}$, connected
with $F\,$(the gravitational-dilatonic one) and $m_{\psi }$ connected with $%
\psi $ (the Polyakov contribution). It is just $m_{F}\neq 0$ which was
implied in \cite{pl}, while the total sum $m_{F}+m_{\psi }=0$ (massless
shell).}. If $q>0$ and the shell is placed at $\phi _{B}$ such that $\exp
(-2\phi _{B})=\exp (-2\phi _{+})(1+\frac{\left| \tilde{D}\right| }{q})$, the
boundary and ordinary quantum corrections mutually cancel.

\section{Energy, ADM\ mass and choice of background}

From the physical viewpoint, the perfect shell considered in the previous
section realizes microcanonical boundary conditions that fixed the energy
(see, cf. \cite{york85}). Meanwhile, another physically relevant type of
conditions demands fixing the temperature rather than the energy, thus
defining the canonical ensemble. This case is also discussed below. In so
doing, the correct definition of thermal quantities, such as the energy,
heat capacity, etc., can be obtained with the help of the Euclidean action
formalism, with account for the finiteness of the system that, in
particular, needs specifying the set of boundary data. Generalizing
expressions for classical gravitation-dilaton systems \cite{action}, one can
write down the energy of the quantum-corrected one as \cite{found}

\begin{equation}
E_{gd}=-\frac{1}{\pi }\left( \frac{d\tilde{F}}{dl}\right) _{B}=-\frac{%
\lambda }{\pi }\left( \frac{d\tilde{F}}{d\phi }\frac{\sqrt{f}}{z^{\prime }}%
\right) _{B}\equiv -2T_{0}\left( \frac{d\tilde{F}}{d\phi }\sqrt{f}\frac{%
\partial \phi }{\partial z}\right) _{B}\text{.}  \label{e}
\end{equation}

For exactly solvable models (see eq. (2.7) of Ref. \cite{exact}) 
\begin{equation}
\frac{dz}{d\phi }=\tilde{F}^{\prime }\frac{\exp (2\phi )}{2}\text{.}
\label{zf}
\end{equation}
Here the common factor in the right hand side (\ref{zf}) is chosen, for the
models (\ref{fv}), to give $z=-\phi +const$ (linear dilaton vacuum) at the
right infinity, where spacetime is flat. Thus, for exactly solvable models
we have 
\begin{equation}
E_{gd}=-4T_{0}\exp (-2\phi _{B})\sqrt{f_{B}}\text{.}  \label{esen}
\end{equation}

In general, the energy $E$ is measured with respect to some background whose
contribution $E_{0}$ is to be subtracted from $E_{gd}$, so $E=E_{gd}-E_{0}$.
In \cite{sol96} two reference points were considered: the classical hot flat
spacetime (which is obtained by putting $\kappa =0$ in the action) and the
black hole configuration with the singular horizon. We adopt another
reference configuration: as a background, we choose the ''quasi-flat''
spacetime which is close to the classical one but differs from it due to the
presence of the terms with $\kappa $ in (\ref{fv}). Let me remind the reader
that the parameter $\kappa $ enters independently both the
Polyakov-Liouville action and the definition of the action coefficients (\ref
{fv}). In the second case it was motivated by the demand to construct
exactly solvable models but now we relaxed that condition. We discard the
first contribution but retain the second one. To avoid confusion, one may
replace $\kappa $ in (\ref{fv}) by another small parameter $\tau $,
effecting the functional form of these coefficients and put $\tau =\kappa $
after calculations. Physically, this means that our reference state is pure
classical in the sense that not any quantum backreaction is present, but the
functional form of the gravitation-dilaton action is the same as for the
quantum-corrected configuration.

To find $E_{0}$, we have to solve field equations (\ref{vx}), (\ref{uf}) for
these potentials without the contribution of $\psi $, so tilted quantities
should be replaced by usual ones. Then, with terms of the order $\kappa ^{2}$
and higher neglected, we find 
\begin{equation}
E_{0}=-2T_{0}\left( \sqrt{f}\frac{\partial F}{\partial \phi }\frac{\partial
\phi }{\partial z}\right) _{quasiflat}\text{,}
\end{equation}
\begin{equation}
f=1-\kappa \frac{c}{2}\phi \exp (2\phi )+\frac{\kappa }{2}\exp (2\phi )(q-1)%
\text{.}
\end{equation}
\begin{equation}
\frac{\partial z}{\partial \phi }=-1-\kappa \frac{c}{4}\exp (2\phi )\text{.}
\end{equation}
Asymptotically, for large $\left| \phi \right| $, $\phi <0$, we obtain 
\begin{equation}
E_{0}=-4T_{0}\exp (-2\phi )+T_{0}\kappa c\phi +T_{0}\kappa (\frac{c}{2}+b)%
\text{.}  \label{e0}
\end{equation}
Now the quantity $E_{g}-E_{0}$ can be identified with the ADM mass and we
have, after asymptotic expansion of $E$, 
\begin{equation}
E_{g}-E_{0}=M_{BH}+M_{th}+M_{0}\text{,}  \label{mt0}
\end{equation}
\begin{equation}
M_{BH}=2T_{0}[\exp (-2\phi _{+})-\kappa \frac{c}{2}\phi _{+}]\text{, }%
M_{th}=2T_{0}\kappa (\phi _{+}-\phi )\text{, }M_{0}=\kappa T_{0}\text{.}
\label{mt1}
\end{equation}

Here the term $M_{BH}$ does not depend on $\phi $ and should be identified
with the mass of a black hole itself. The quantity $M_{th}$ represents the
contribution of thermal gas at the temperature $T_{0}$. Remarkably, the
coefficients $b$ and $c$, that characterize the model, are absorbed by these
general definitions. One can say that not only for the RST\ model \cite
{sol96} and even not only for a more general exactly solvable model, but in
the general case for the family (\ref{fv}), quantum corrections to the
universal form (\ref{mt0}) vanish.

In a similar way, the total entropy $S_{tot}=S_{BH}+S_{th}$, where the
entropy of the black hole itself

\begin{equation}
S_{BH}=2F(\phi _{+})=2[\exp (-2\phi _{+})+b\kappa \phi _{+}]=\frac{M_{BH}}{%
T_{0}}+2\kappa (q-1)\phi _{+}\text{,}
\end{equation}
while the entropy of the thermal gas \cite{sol96}, \cite{exact}, \cite
{action}, \cite{entropy} 
\begin{equation}
S_{th}(T_{0},\phi _{+}-\phi _{B})=4\kappa (\phi _{+}-\phi _{B})\text{.}
\end{equation}

The expression (\ref{mt0}) is valid in an infinite space. Let now a wall be
placed at $\phi =\phi _{B}$. Take $\phi =\phi _{0}$ to the right from the
boundary and consider the region between the boundary and infinity. Taking
into account (\ref{dz}) - (\ref{d}), (\ref{ga}) and expanding the expression
for the energy for large negative $\phi _{0}$, where spacetime approaches
its Minkowski limit, we obtain after simple calculations: 
\begin{equation}
E_{g}=4\kappa (T-T_{0})-4T_{0}\exp (-2\phi )+2T_{0}k+T_{0}\kappa
[q+2D+(c-2+2D)\phi ]\text{,}  \label{egg}
\end{equation}
\begin{equation}
M_{tot}=M_{BH}+M_{th}(T_{0}\text{, }\phi _{+}-\phi _{B})+M_{th}(T\text{, }%
\phi _{B}-\phi _{0})+M_{0}\text{, }  \label{mt}
\end{equation}

\begin{equation}
M_{th}(T\text{, }\phi _{1}-\phi _{2})=\frac{\pi }{6\lambda }T^{2}(\phi
_{1}-\phi _{2})=2\kappa \frac{T^{2}}{T_{0}}(\phi _{1}-\phi _{2})\text{.}
\end{equation}
The formula (\ref{mt}) generalizes (\ref{mt0}) in a natural way: it includes
the contribution of the thermal gas with two different temperatures - $T_{0}$
between the horizon and the wall and $T$ between the wall and the point of
observation. More surprisingly, vacuum polarization in the Boulware state
with $T=0$ (when quantum stresses do not vanish) does not give corrections
at all, thus the only contribution of the state outside comes due to thermal
excitations, if the Boulware state is heated to some temperature $T$. This
fact can be attributed to the change of the effective coupling between the
curvature and dilaton: the quantity $\tilde{F}$ changes to $\tilde{F}+\kappa 
\frac{\gamma }{\lambda }y$ in such a way that the first term in (\ref{egg})
cancels the vacuum contribution.

In Sec. III we obtained that, if we want the shell to maintain thermal
equilibrium inside and the Boulware state outside, it cannot be placed too
closely to the horizon. Now, the general formulas for the energy obtained
above, enable us to give a rather simple physical interpretation to the
corresponding restriction on the position of the shell. Let the quantum
state be in the Boulware state ($T=0$, $\left| \tilde{B}\right| =\kappa $, $%
D=1$) outside. To elucidate the role of different terms containing the
parameter $\kappa $, we consider the case, when the restriction under
discussion is obtained exactly (\ref{res}). Taking into account the explicit
expression for the action coefficients (\ref{fv}), the conditions of
solvability (\ref{esq}), restoring explicitly the factor $\frac{\lambda }{%
\pi }=2T_{0}$, so that the mass of the shell $M_{shell}=2T_{0}m$, and using
the expression for the total mass of black hole plus thermal radiation $%
M_{tot}$ (\ref{mt}), we obtain 
\begin{equation}
M_{tot}^{\prime }>M_{tot}+M_{shell}^{\prime }\text{.}  \label{hhb}
\end{equation}

Here $M_{shell}^{\prime }=M_{shell}+3M_{0}$, and $M_{tot}^{\prime }$ $\equiv
2T_{0}[\exp (-2\phi _{B})-\kappa \frac{c}{2}\phi _{B}]$ represents the mass
of a black hole which would form, if thermal radiation completely collapsed,
producing a new black hole with the horizon at $\phi _{+}^{\prime }=\phi
_{B} $. This horizon would coincide with the radius of the corresponding 1+1
''relativistic star'', the quantum field outside being in the Boulware state.

\section{Canonical ensemble and heat capacity}

In general, for self-gravitating systems the conditions of stability can be
different for different types of thermal ensembles. As far as our
gravitation-dilaton system is concerned, the stability of the microcanonical
ensemble (without account for boundary corrections) follows immediately from
eq. (12)\ of Ref. \cite{phase} (case $a=0$ in their notations). The case of
the canonical ensemble is much more subtle since it demands simultaneous
careful account for the finiteness of the system and quantum backreaction.
Let us discuss this issue in more detail.

The canonical ensemble is defined by the value of temperature and, possibly,
some other parameters, which for self-gravitating systems are fixed on the
boundary \cite{york86}. According to the Tolman relation, 
\begin{equation}
T_{B}=\frac{T_{H}}{\sqrt{f_{B}}}\text{,}  \label{tol}
\end{equation}
where $T_{B}$ is the local temperature on the boundary. For the system under
discussion the value of the dilaton $\phi _{B}$ is also fixed. The region,
external with respect to the boundary, is now discarded and replaced by a
heat bath, so there is no sense in speaking about boundary corrections to
the Hawking temperature. Nevertheless, the finiteness of the system reveals
itself, as we will see below, in the dependence of the horizon radius $\phi
_{+}$ and thermodynamic characteristics on the boundary data.

First, discuss briefly the exactly solvable case. Then $T_{H}=T_{0}=const$ 
\cite{exact}. Then it follows from (\ref{ff}) that, for given $T_{B}$ and $%
\phi _{B}$, there is also one root $\tilde{F}_{+}$. If the function $\tilde{F%
}(\phi )$ is monotonic (for example, this happens to the BPP model), there
is only one branch and one value $\phi _{+}$. In general, this function can
have minima and maxima. For example, in the RST model there are two branches
of solutions: the upper branch $\phi _{s}<\phi <\infty $ and the low branch $%
-\infty <\phi <\phi _{s}$, where $\phi _{s}$ corresponds to the singularity 
\cite{sol96}.

The heat capacity can be found from (\ref{esen}), (\ref{tol}) (now the term $%
E_{0}$ does not contribute and can be omitted): 
\begin{equation}
C=\frac{dE}{dT_{B}}=4\exp (-2\phi _{B})f_{B}=4\exp (-2\phi _{B})\frac{%
T_{0}^{2}}{T_{B}^{2}}>0\text{.}  \label{ces}
\end{equation}
Of main interest is the region between the horizon and Minkowski spacetime
at infinity, with $-\infty <\phi <\phi _{+}$. Then it follows from explicit
expressions (\ref{fv}), (\ref{ff}) that in this region always $0\leq f<1$.
Therefore, if $T_{B}>T_{0}$, there is 1 stable root. If $T_{B}<T_{0}$, there
are no roots at all. Thus, if the equilibrium is possible, the system is
always locally stable.

Consider now the case with generic coefficients $b$, $c$ within the
perturbative approach with respect to $\kappa $. Now, in contrast to the
exactly solvable case, quantum corrections to the Hawking temperature do not
vanish and an interesting overlap between quantum and boundary effects
appears. Remembering (\ref{t}) and differentiating the relevant quantities,
we obtain 
\begin{equation}
\frac{\partial E}{\partial \phi _{+}}=-2\frac{T_{0}}{\sqrt{f}}\frac{\partial
f}{\partial \phi _{+}}\exp (-2\phi _{B})[1-\kappa \frac{q}{2}\exp (2\phi
_{B})]\text{.}
\end{equation}
\begin{equation}
\frac{\partial T}{\partial \phi _{+}}=-\frac{T_{0}}{2f\sqrt{f}}\{\frac{%
\partial f}{\partial \phi _{+}}-2f\kappa q\exp (2\phi _{+})]\text{.}
\end{equation}
\begin{equation}
C=4\exp (-2\phi _{B})\frac{f^{\prime }[1-\kappa \frac{q}{2}\exp (2\phi _{B})]%
}{f^{\prime }-2f\kappa q\exp (2\phi _{+})}\text{,}
\end{equation}
where $f^{\prime }=\frac{df}{d\phi _{+}}$. Let $\phi _{B}\rightarrow -\infty 
$, $\kappa \rightarrow 0$, then in the main approximation $\frac{\partial f}{%
\partial \phi _{+}}=2\exp (2\phi -2\phi _{+})$.

Writing $\frac{T}{T_{0}}\equiv 1+\alpha $ with small, but non-zero $\alpha $%
, we obtain from (\ref{tol}) the equation 
\begin{equation}
r^{2}-2rr_{0}+s=0\text{, }r\equiv \exp (-2\phi _{+})>0\text{, }r_{0}\equiv
\exp (-2\phi _{B})\alpha \text{, }s\equiv \kappa q\exp (-2\phi _{B})\text{,}
\label{rs}
\end{equation}
where parameters $r_{0}$ and $s$, constructed as the products of small and
big quantities, are in general finite. Writing the solution of the quadratic
equation as $r_{\pm }=r_{0}\pm \sqrt{r_{0}^{2}-s}$, we find 
\begin{equation}
C=4\exp (-2\phi _{B})\frac{r^{2}}{r^{2}-s}=\pm 2\exp (-2\phi _{B})\frac{%
r_{\pm }}{\sqrt{r_{0}^{2}-s}}\text{,}  \label{c}
\end{equation}

where we took into account that $\kappa \phi _{B}\exp (2\phi _{B})\ll 1$.
Thus, only the root $r_{+}$ can correspond to the stable equilibrium. If $%
r_{0}^{2}<s$, there are no black hole solutions in thermal equilibrium at
all, so the ground state lies in the same topological sector as the flat
spacetime. Let $r_{0}^{2}>s$ and discuss now particular cases.

1) $r_{0}>0$, $s>0$. Then $r_{+}>0$, $r_{-}>0$. There are 2 roots: $r_{+}$
is stable, $r_{-}$ is unstable.

2) $r_{0}<0$, $s>0$. $r_{+}<0$, $r_{-}<0$. There are no positive roots at
all.

3) $r_{0}>0$, $s<0$. $r_{+}>0$, $r_{-}<0$. 1 stable root $r_{+}$.

4) $r_{0}<0$, $s<0$. $r_{+}>0$, $r_{-}<0$. 1 stable root $r_{+}$.

Thus in cases 1), 3), 4) we have the locally stable black hole solution.
Pure classical consideration ($\kappa =s=0$) would give only 1 stable root $%
r_{cl}=2r_{0}$, and $C=4\exp (-2\phi _{B})>0$, $C\rightarrow \infty $ for $%
\phi _{B}\rightarrow -\infty $. In case 1) the root $r_{+}<r_{cl}$, whereas
in case 3) $r_{+}>r_{cl}$. According to (\ref{z}), the spatial coordinate $z$
grows, when $\phi $ diminishes. Therefore, in case 1) quantum corrections
decrease the horizon radius, whereas in case 3) it slightly increases, as
compared to the classical case. Case 4)\ is pure quantum and does not exist
in the classical domain. Indeed, for $\kappa $ so that $s\ll r_{0}^{2}$, the
root $r_{+}\simeq \frac{\left| s\right| }{2\left| r_{0}\right| }=\frac{%
\kappa }{2}\left| \frac{q}{\alpha }\right| $ is proportional to the quantum
parameter $\kappa $.

The Euclidean action $I=\beta E-S_{tot}$, in the main approximation ($\kappa
\rightarrow 0$, $\phi _{B}\rightarrow -\infty $) $I=$ $-2\kappa (\phi
_{+}-\phi _{B})<0$. Thus, if a black hole solution exists, it is a favorable
phase and is stable not only locally, but also globally. For $q=s=0$, $%
r_{+}=2r_{0}$ and we return to a exactly solvable model.

It is instructive to discuss case 1) in more detail to reveal the role of
the finiteness of a system in the issue of stability. Let us suppose, for a
moment, that we proceed in an infinite space from the very beginning and,
substituting $f=1$ at infinity in (\ref{tol}), identify $T=T_{H}$. The
formula (\ref{t}) can be rewritten, in the main approximation with respect
to $\kappa $, as 
\begin{equation}
T_{H}=T_{0}(1+\kappa \frac{T_{0}\delta }{M_{BH}})\text{, }\delta =q\text{.}
\label{qd}
\end{equation}
Then, direct differentiation gives us 
\begin{equation}
C=\left( \frac{dT_{H}}{dM_{BH}}\right) ^{-1}=-\frac{M_{BH}^{2}}{\kappa
qT_{0}^{2}}=-\frac{4\exp (-4\phi _{B})}{\kappa q}\text{.}  \label{cw}
\end{equation}

It would seem that the sign of the coefficient $\delta $ is crucial in that
it determines the sign of the heat capacity and stability or instability of
the canonical ensemble. In particular, direct application of (\ref{cw}) or
eq. (18) of \cite{pl} to the CGHS model (for which $q>0$) leads to the
conclusion about instability \cite{pos}.

However, such consideration does not exhaust all possible solutions for case
1). Formally, the quantity (\ref{cw}) can be obtained from the first
equality in (\ref{c}), if the term $r^{2}=\exp (-4\phi _{+})\,$is finite,
whereas $s=\kappa q\exp (-2\phi _{B})$ grows for $\phi _{B}\rightarrow
-\infty $. Then $r^{2}$ can be neglected in the denominator. Meanwhile, the
point is that in case 1) there exist {\it two} different roots. When a size
of a system is large ($-\phi _{B}\gg 1$), in this limit $r_{0}^{2}\gg s$.
Correspondingly, $r_{+}\simeq 2r_{0}$, $r_{-}\simeq s/2r_{0}=\frac{\kappa q}{%
2\alpha }$. Thus, the root $r_{-}$ does not depend on $\phi _{B}$ in the
limit of large $\left| \phi _{B}\right| $, when the boundary is placed in
the nearly flat region, whereas the root $r_{+}$ itself grows, as it follows
from (\ref{rs}). Therefore, the inequality $r^{2}\ll s$ is valid only for $%
r_{-}$, but not for $r_{+}$. As a result, the prediction of the negative
heat capacity on the basis of eq. (\ref{cw}) refers to the root $r_{-}$ only
which is indeed unstable. However, for the root $r_{+}$ the horizon radius $%
\phi _{+}=\phi _{B}$ + const approaches infinity in the same manner as $\phi
_{B}$ so does. Therefore, eq. (\ref{cw}), derivation of which tacitly
assumes that $\left| \phi _{+}\right| $ is finite, while $\left| \phi
_{B}\right| \gg 1$, does not work now. One is forced to use eq. (\ref{c}),
not discarding the $r^{2}$ term in the denominator, whence it is seen that
the root is indeed {\it stable}. In the limit under discussion the heat
capacity $C=4\exp (-2\phi _{B})$ looks very much like in the case of exactly
solvable models (\ref{ces}) in spite of the fact that now $q\neq 0$. In the
limit $\phi _{B}\rightarrow -\infty $ the heat capacity diverges, but for
any large but finite $\phi _{B}$ it is finite and positive.

To great extent, the situation resembles the one for Schwarzschild black
holes in the canonical ensemble \cite{york86}. Naive application of the
formula for the Hawking temperature $T=(8\pi M)^{-1}$ would give the heat
capacity $C=\frac{dM}{dT}=-8\pi M^{2}<0$ with the conclusion about
instability. However, thorough treatment showed that, for a given physical
temperature $T$ on the boundary, there exists two different positions of the
horizon as roots $r_{+}$ and $r_{-}$ of eq. (\ref{tol}). The light root $%
r_{-}$ $<r_{+}$ has the horizon radius $2M=(4\pi T)^{-1}$, when the radius
of the boundary $r_{B}\rightarrow \infty $, and for it the calculation of
the heat capacity in an infinite space is justified with the conclusion
about instability of the solution. But the heavy root $r_{+}$ itself tends
to $r_{B}$, when $r_{B}\rightarrow \infty $ and one cannot apply to it
formulas in an infinite space, ignoring the boundary. Careful treatment
shows that for this root the ensemble is stable \cite{york86}. In our 2D
system also it is the ''heavy'' root which is stable, whereas the ''light''
one is unstable. In both cases (for our system and for Schwarzschild black
holes) one loses the heavy solution, which is most important physically, if
the presence of the boundary is ignored. On the other hand, the difference
between these two situations lies in that this effect manifests itself in 
\cite{york86} on the pure classical level, while for our case it is relevant
only if quantum backreaction is taken into account.

\section{Summary}

We considered a generic string-inspired gravitation-dilaton model, that are
characterized by numeric parameters, for particular values of which a model
becomes exactly solvable. Quantum corrections to the Hawking temperature of
a black hole in an infinite space are found. We analyzed also, how the
presence of a finite size cavity affects thermal properties of a black hole.
Two types of different boundary conditions are considered - microcanonical
and canonical ones. In the first case there is a perfectly reflecting shell
that fixes the energy inside. We found, how vacuum polarization outside the
shell affects thermodynamics of a black hole inside the shell and calculated
the corrections to the Hawking temperature due to the shell. As a
by-product, it turned out that the shell cannot be placed as near to the
horizon as one likes. In the second case the outer space is removed and is
replaced by the heat bath. It is shown that the canonical ensemble is
well-defined and stable in the wide region of parameters. Accounting for the
finiteness of the system is important to the extent that in some cases it
alters the conclusion about instability (typical of consideration in an
infinite space) and gives stable solutions. In so doing, quantum
backreaction is also important. In particular, the type of solutions is
found which exists only due such a backreaction.

The indirect dependence of black hole thermodynamics on vacuum polarization
outside a shell should also be relevant for 4D black holes \cite{boulw}. In
this respect 2D dilaton gravity revealed itself one more time as a clear and
simplified tool for understanding overlap between quantum theory and
gravitation that occurs in our real world.

\section{Acknowledgment}

I wish to thank Daniel Grumiller, Dmitri Vassilevich and Wolfgang Kummer for
intensive correspondence which stimulated the appearance of this article.





%
%

%
%

\end{document}